\documentclass[twocolumn]{aastex631}

\begin{document}

\title{Discovery of a radio megahalo in the cluster PLCKG287.0+32.9 using the uGMRT}

\correspondingauthor{Sameer Salunkhe}
\email{ssalunkhe@ncra.tifr.res.in}

\author[0000-0002-0666-2326]{Sameer Salunkhe}
\affiliation{National Centre for Radio Astrophysics, Tata Institute of Fundamental Research, Pune 411007, India}

\author[0009-0002-0373-570X]{Ramananda Santra}
\affiliation{National Centre for Radio Astrophysics, Tata Institute of Fundamental Research, Pune 411007, India}

\affiliation{INAF - IRA, Via Gobetti 101, I-40129 Bologna, Italy}

\author[0000-0003-1449-3718]{Ruta Kale}
\affiliation{National Centre for Radio Astrophysics, Tata Institute of Fundamental Research, Pune 411007, India}



\begin{abstract}

We report the discovery of a radio megahalo in the merging cluster PLCKG287.0+32.9, based on upgraded Giant Metrewave Radio telescope (uGMRT) observations at frequencies 300-850 MHz. The sensitive radio observations provide a new window to study the complex physics occurring in this system. Apart from significant detections of the known diffuse radio emission in the cluster, we detect the central diffuse emission to a much larger extent of $\sim$ 3.2 Mpc, reaching the R$_{500}$ of the cluster. The radial surface brightness profile shows a distinct flattening beyond $\sim$0.5R$_{500}$, dividing the emission into inner and outer components. This outer envelope shows a steep spectral index ($\lesssim$ -1.5) and, emissivity $\sim 20$ times lower than the inner component, confirming the megahalo characteristics. The radial profile of the spectral index also distinguishes the steep spectrum megahalo emission. Our observational results align with recent numerical simulations, showing megahalo emission oriented along the merger axis and the re-acceleration of electrons driven by late-stage merger-induced turbulence. This is the first detection of a radio megahalo at a frequency higher than the LOFAR 144 MHz, opening the possibilities for more discoveries and spectral studies to understand their origin. 
\end{abstract}




\section{Introduction} \label{sec:intro}

The observational footprints of galaxy cluster merger events are seen in some clusters via Mpc-scale diffuse radio emission, not associated with individual galaxies, broadly classified as radio halos and radio relics \citep[for reviews;][]{2019SSRv..215...16V, 2023JApA...44...38P}. Radio relics are highly polarised (up to 50\%), steep spectrum sources ($\alpha$\footnote{S $\propto \nu^{\alpha}$, where $\alpha$ is the spectral index and S is flux density} $<-1$), located at the cluster peripheries, and are thought to be the tracers of the outgoing merger shock waves \citep[e.g.,][]{2008A&A...486..347G, 2012MNRAS.426.1204K, 2022A&A...659A.146D}. Radio halos are located in the central regions of galaxy clusters and tend to trace the X-ray-emitting thermal gas \citep[e.g.,][]{2009A&A...507.1257G, 2022MNRAS.514.5969K, 2024ApJ...976...66S}. Currently, the favored scenario for the generation of the radio halo involves the re-acceleration of the pre-existing relativistic electrons, due to merger-induced turbulence \citep[turbulent re-acceleration;][]{2007MNRAS.378..245B, 2016MNRAS.458.2584B, Fujita_2015}. Even relaxed cool core clusters are found to host central diffuse radio emission surrounding the brightest cluster galaxy (BCG), and is referred to as a mini-halo (see \citealt{giacintucci14,giacintucci17,giacintucci19,riseley23,riseley24}). The origin of the mini-halo emission is often attributed to either the hadronic model or re-acceleration due to turbulence associated with the sloshing of the cluster core \citep[e.g,][]{2004A&A...413...17P, 2013ApJ...762...78Z, 2015ApJ...801..146Z}. Despite several observational and theoretical studies, the detailed physics behind the origin of these sources is not fully understood \citep[e.g.,][]{2014IJMPD..2330007B}.

With the advent of low-frequency radio observations, new low surface brightness and steep spectrum radio sources have been detected, including large-scale bridges between pre-merging clusters \citep[e.g.,][]{2019Sci...364..981G, 2020MNRAS.499L..11B, 2024A&A...682A.105P, 2024A&A...691A..99P}, fossil plasma sources \citep{2020A&A...634A...4M, 2022A&A...664A.186S}, mini-halos that extend beyond the cool cores and embedded within radio halos \citep[e.g.,][]{2019MNRAS.486L..80K, 2019A&A...622A..24S, 2022MNRAS.512.4210R, 2024A&A...686A..82B, 2024A&A...692A..12V} as well as three-component halos \citep{2023A&A...678A.133B}. Particularly noteworthy, \citet{2022Natur.609..911C} identified a new type of diffuse source in four clusters, termed megahalo, which envelops classical radio halo. Megahalos extend up to scales approximately equal to R$_{500}$ (the radius at which the mean mass density is 500 times the critical density of the universe at that redshift) of the cluster and validate the extension of relativistic electrons and magnetic fields well beyond the radius of radio halos. These megahalos display a shallower surface brightness radial profile, a very steep spectrum (with a spectral index exceeding -1.6 between approximately 50 and 150 MHz), and a lower emissivity (20 - 25 times), compared to their embedded radio halos suggesting distinct physical conditions in the outer regions of clusters compared to those within radio halos \citep{doi:10.1126/sciadv.abq7623,2022Natur.609..911C}. Thus, uncovering the origin of the megahalo emission is a key challenge in advancing our understanding of the evolution of the non-thermal population in galaxy clusters.

Radio megahalos, in general, are difficult to study owing to their low surface brightness and steep radio spectrum, making them faint at high frequencies. They are difficult to detect in nearby clusters, as the structures will be resolved out and at high redshift the structure will be difficult to detect due to cosmological dimming \citep{2022Natur.609..911C}. Therefore, for the search of megahalos, the massive clusters with moderate redshift can serve as the best candidates if observed with low frequencies. One such ideal test bed is the cluster PLCKG287.0+32.9 (hereafter PLCKG287). PLCKG287 is a highly X-ray luminous (L$_{\rm 0.1-2.4 keV}$ = 1.72$\pm$0.01$\times$ 10$^{45}$ erg.s$^{-1}$ within R$_{500}$ = 1541 kpc) and massive (M$_{500}$ = 1.47$\pm$0.04$\times$ 10$^{15}$ M$_{\odot}$) cluster situated at a redshift of 0.39 \citep{2011ApJ...736L...8B, 2016A&A...594A..27P}. The cluster is reported to host a pair of gigantic relics, a radio halo, and many filamentary emissions at multiple frequencies using the GMRT (Giant Metrewave Radio Telescope), VLA (Very Large Array), and MWA (Murchison Widefield Array) \citep{2011ApJ...736L...8B, 2014ApJ...785....1B, 2017MNRAS.467..936G}.

This letter presents the discovery of a radio megahalo from extensive radio observations conducted on the galaxy clusters PLCKG287, using the uGMRT. We adopt a Lambda cold dark matter cosmology with $\rm H_0=70$~km~s$^{-1}$~Mpc$^{-1}$,  $\Omega_\Lambda = 0.7$ and $\Omega_m = 0.3$. In this framework, $1\arcsec$ corresponds to $5.29$ kpc at the cluster's redshift ($z = 0.39$), and the luminosity distance ($\rm D_{\rm L}$) is $2108.2$~Mpc.

\section{Observations and data analysis}\label{sec:data}

The galaxy cluster PLCKG287 was observed with the uGMRT (observation code: 45\_099) in band 3 and band 4. The observations at both frequency bands were done using the GMRT Wideband Backend (GWB) covering a frequency range of 300-500 MHz for band 3 and 550-950 MHz for band 4. However, only 550-850 MHz bandwidth is used for band 4 imaging. We used real-time online RFI filtering in GWB observation, to mitigate the broadband RFI \citep{buch_2022JAI....1150008B, buch_2023JApA...44...37B}. 

Analyzing data at lower frequencies poses challenges due to the influence of ionospheric effects, which limit the dynamic range of the image. To address this, we employ the Source Peeling and Atmospheric Modeling ({\texttt{ SPAM}, \citealt{Intema_2009A&A...501.1185I, Intema_2017A&A...598A..78I}) pipeline for uGMRT data analysis. This comprehensive pipeline encompasses standard data reduction procedures and tackles the direction-dependent calibration required to account for fluctuations in visibility amplitude and phase, stemming from antenna beam patterns and the ionosphere. This direction-dependent calibration leverages bright sources within the primary beam (for more details, refer to \citealt{Intema_2009A&A...501.1185I, Intema_2017A&A...598A..78I}). Initially, the wide-band uGMRT data were split into six narrow bands, each of width $\sim33$~MHz for band 3 and $\sim50$~MHz for band 4. Both sets of six narrow bands were then separately calibrated for flux and band shape using the flux calibrators 3C147 at band 3 and 3C286 at band 4. The absolute flux density scale was set according to \citet{2012MNRAS.423L..30S}. The images for all the narrow bands with direction-dependent calibrations were produced in \texttt{SPAM} for both uGMRT bands. Further, the  \texttt{wsclean} package \citep{offringa_2014MNRAS.444..606O} is employed to produce the final wide-band images by combining all six sub-bands as well as the images with different resolutions. All the radio images are primary beam corrected. 

The presence of several discrete sources makes it difficult to measure the integrated flux density of the diffuse emission accurately. Therefore, we have subtracted the point sources for both the uGMRT frequencies, using their models created by applying a \textit{uv}-cut of \textgreater 4 k$\lambda$. The baseline-restricted model component corresponding to the ``discrete'' sources, was Fourier transformed and subtracted from the observed data using \texttt{uvsub} in \texttt{CASA}. After subtraction, the visibilities were imaged using \texttt{wsclean} with \texttt{multi-scale} option to highlight the extended emission at different resolutions. We tested the consistency of the subtraction process by comparing the flux density of the central diffuse emission with the value obtained by algebraically subtracting the flux density of the embedded sources from the total flux density (diffuse emission + compact sources). The results showed good agreement, with the flux density matching within 3\% for GMRT. For further details we refer to \citet{2021A&A...650A..44B}.

\begin{figure}
	
    \includegraphics[width=\columnwidth]{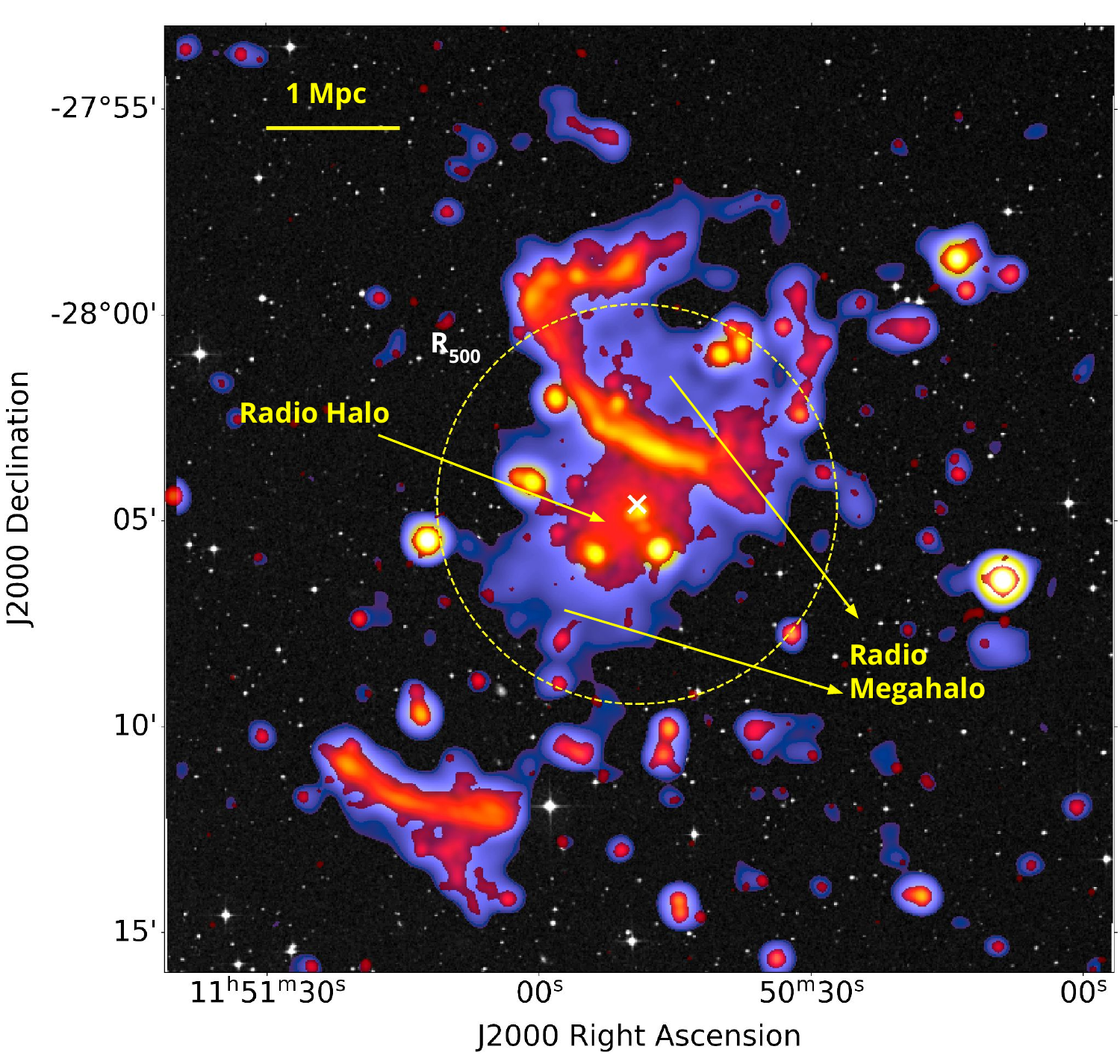}
    \caption{DSS2 i band optical image overlaid with uGMRT band 3 ($\sigma_{\rm rms}=60\mu$Jy beam$^{-1}$, where beam = $30\arcsec \times 30\arcsec$) in violet and band 4 ($\sigma_{\rm rms}=20\mu$Jy beam$^{-1}$, where beam = $8.9\arcsec \times 7.8\arcsec$) in red raster. The yellow dashed circle indicates the R$_{500}$ of the cluster.}     
    \label{megahalo_rgb}

\end{figure}

\begin{figure*}
\centering
\includegraphics[width=8cm, height =8cm]{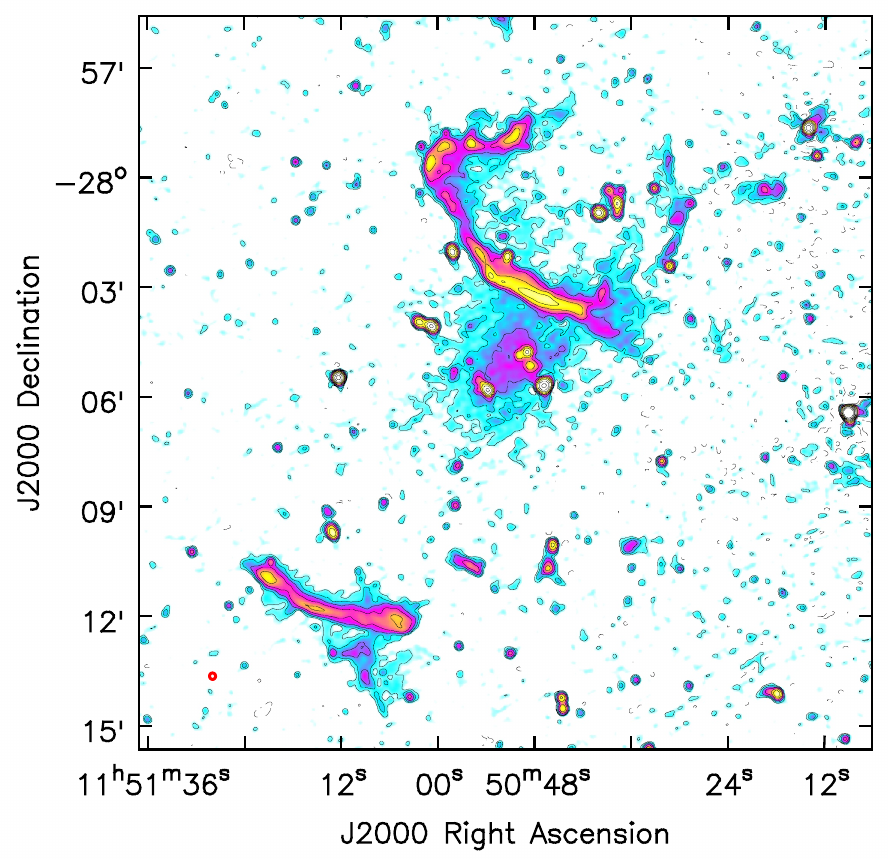}
\includegraphics[width=8cm, height =8cm]{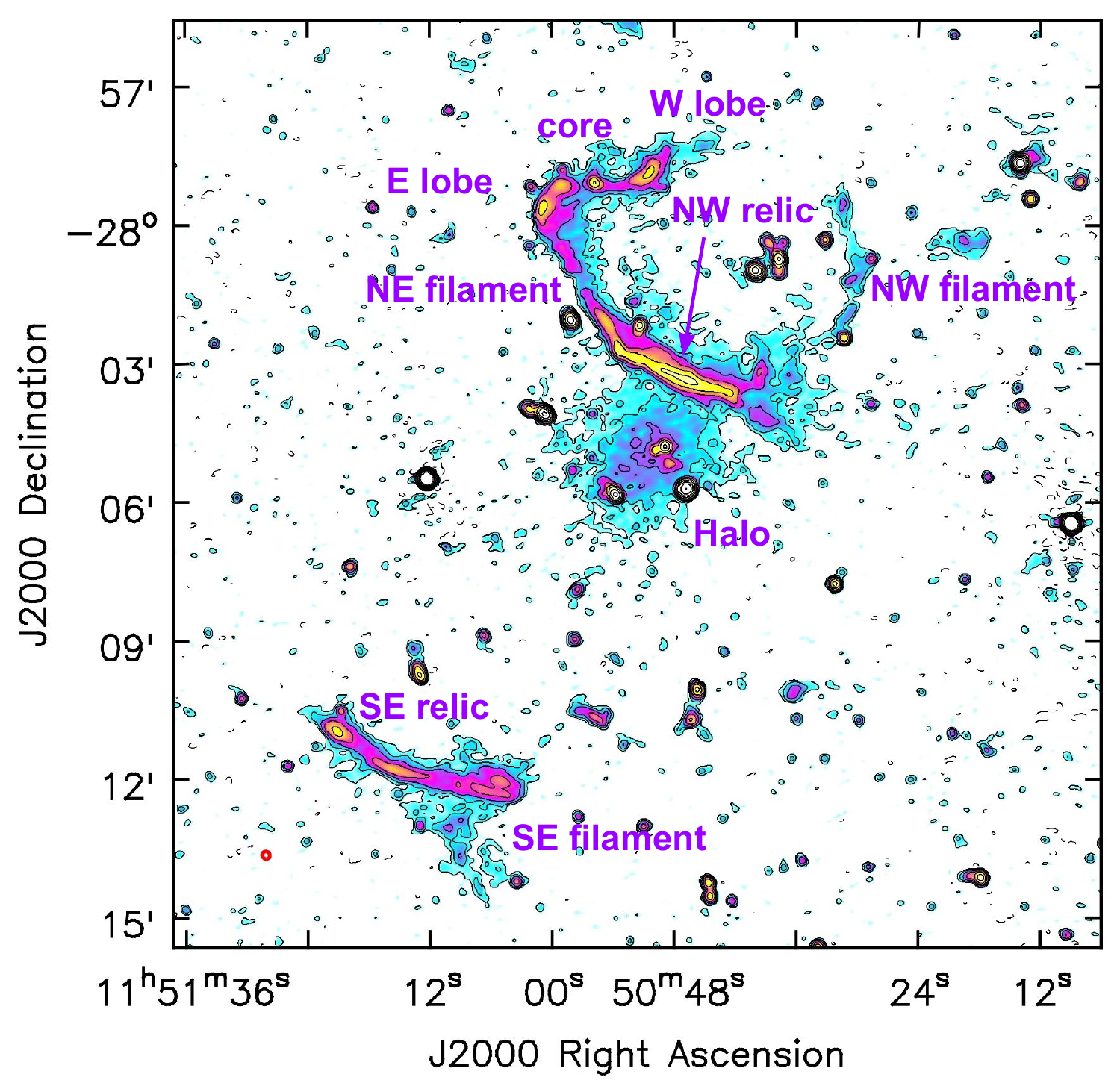}\\
\includegraphics[width=8cm, height =8cm]{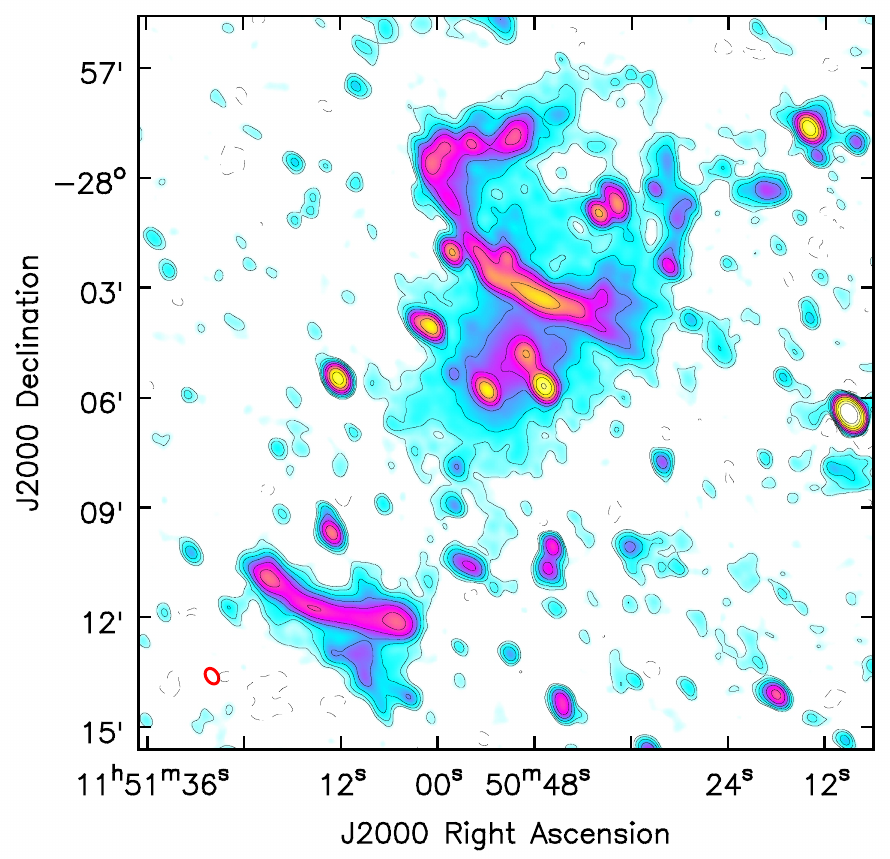}
\includegraphics[width=8cm, height =8cm]{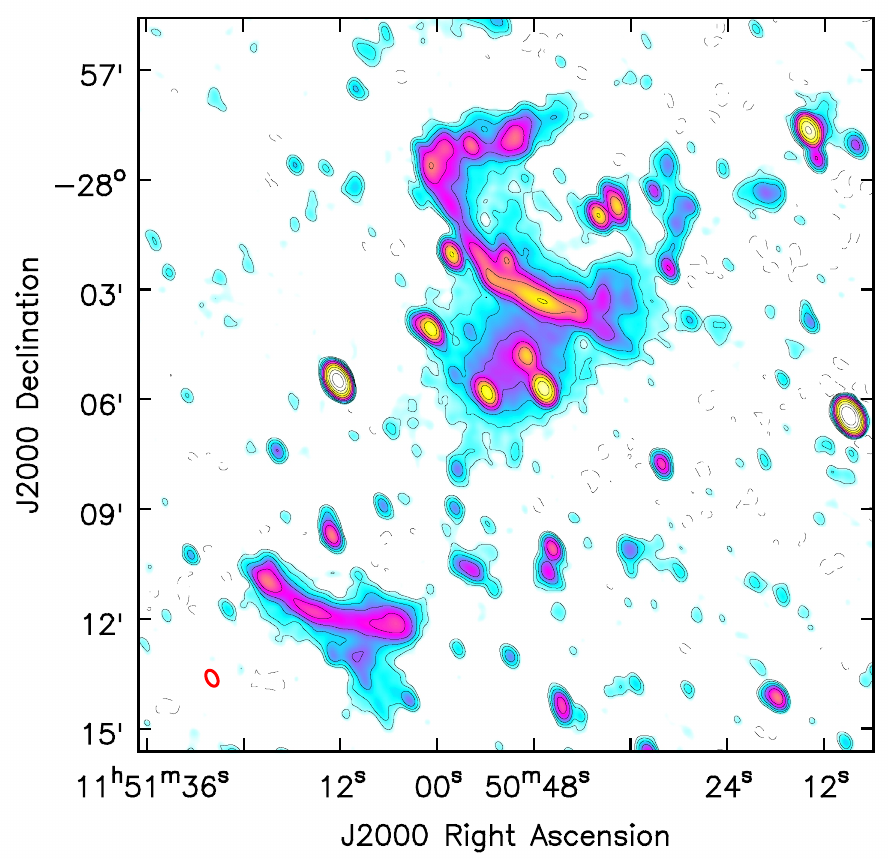}
\caption{\textit{Top left:} uGMRT band 3 high-resolution image (10.5$''$ $\times$ 9.8$''$, PA: 9.3$^{\circ}$ and $\sigma_{\rm rms}$ = 35 $\mu$Jy beam$^{-1}$). \textit{Top right:} uGMRT band 4 high-resolution image (10.4$''$ $\times$ 9.6$''$, PA: 25.9$^{\circ}$, $\sigma_{\rm rms}$ = 25 $\mu$Jy beam$^{-1}$).  Previously observed radio sources are labeled here. \textit{Bottom left:} uGMRT band 3 low-resolution image (27.3$''$ $\times$ 19.4$''$, PA: 33.1$^{\circ}$, $\sigma_{\rm rms}$ = 60$\mu$Jy beam$^{-1}$). \textit{Bottom right:} uGMRT band 4 low-resolution image (27.8$''$ $\times$ 17.5$''$, PA: 27.2$^{\circ}$, $\sigma_{\rm rms}$ = 45$\mu$Jy beam$^{-1}$). Contours level for all images are drawn at [$\pm1,2,4,8,...$]$\times$ 3$\sigma_{\rm rms}$, where $\sigma_{\rm rms}$ is RMS noise of respective image. The red ellipse in the bottom left corner of each panel indicates the beam size of the image in the respective panel. }

\label{Band3_band4}
\end{figure*}

\section{Results} \label{sec:results}

\subsection{Continuum images} \label{subsec: continuum_images}

In Figure~\ref{megahalo_rgb} we show the color composite image of the uGMRT band 3 and band 4, along with the DSS2 optical image of the cluster field. The spectacular extended emission is detected at the cluster center as well as in the periphery. In Figure~\ref{Band3_band4}, we show the resulting Band 3 and Band 4, uGMRT radio continuum images of PLCKG287. To highlight the extended emission, we have made different spatial resolution images at both frequencies, using \texttt{uvtapers}. All the diffuse sources: the halo, two relics, and the other filaments are recovered in our uGMRT images (Figure~\ref{Band3_band4}). The central region of the cluster is permeated by the large-scale low surface brightness diffuse emission, which is the primary focus of this letter. 

As shown in Figure~\ref{megahalo_rgb}, the central diffuse emission is not circularly symmetric, rather it is elongated in the northwest (NW) and southeast (SE) directions, i.e., along the merger direction. The diffuse emission overlaps with the NW relic very smoothly, and extends beyond that. It follows the morphology of the thermal emission from the ICM very well as shown in Figure~\ref{X-ray_overlay}. \cite{2014ApJ...785....1B} had reported a southern extension in the radio halo, with no counterpart of that extension in the X-ray image of the cluster. The total extent of the halo emission is larger than the previously reported size in \citet{2014ApJ...785....1B} at these frequencies. In the 325 and 610 MHz narrow-band images, mainly the innermost part of the halo is detected, while the low surface brightness emission is completely missed very likely because of the poor \texttt{uv-coverage} at short baselines and low sensitivity. From our new wide-band images, the exact Largest Linear Size (LLS) of the central diffuse emission is difficult to measure because of the presence of the extended NW relic. However, the projected LLS of the central diffuse emission is $\sim 3.2$ Mpc, and $\sim 2.9$ Mpc at 400 and 700 MHz respectively, measured from  40$''$ resolution images (Figure~\ref{X-ray_overlay}) where the total diffuse emission is best detected. The central diffuse emission is extended up to the cluster's R$_{500}$ at 400 MHz.

\begin{figure}
    \includegraphics[width=\columnwidth]{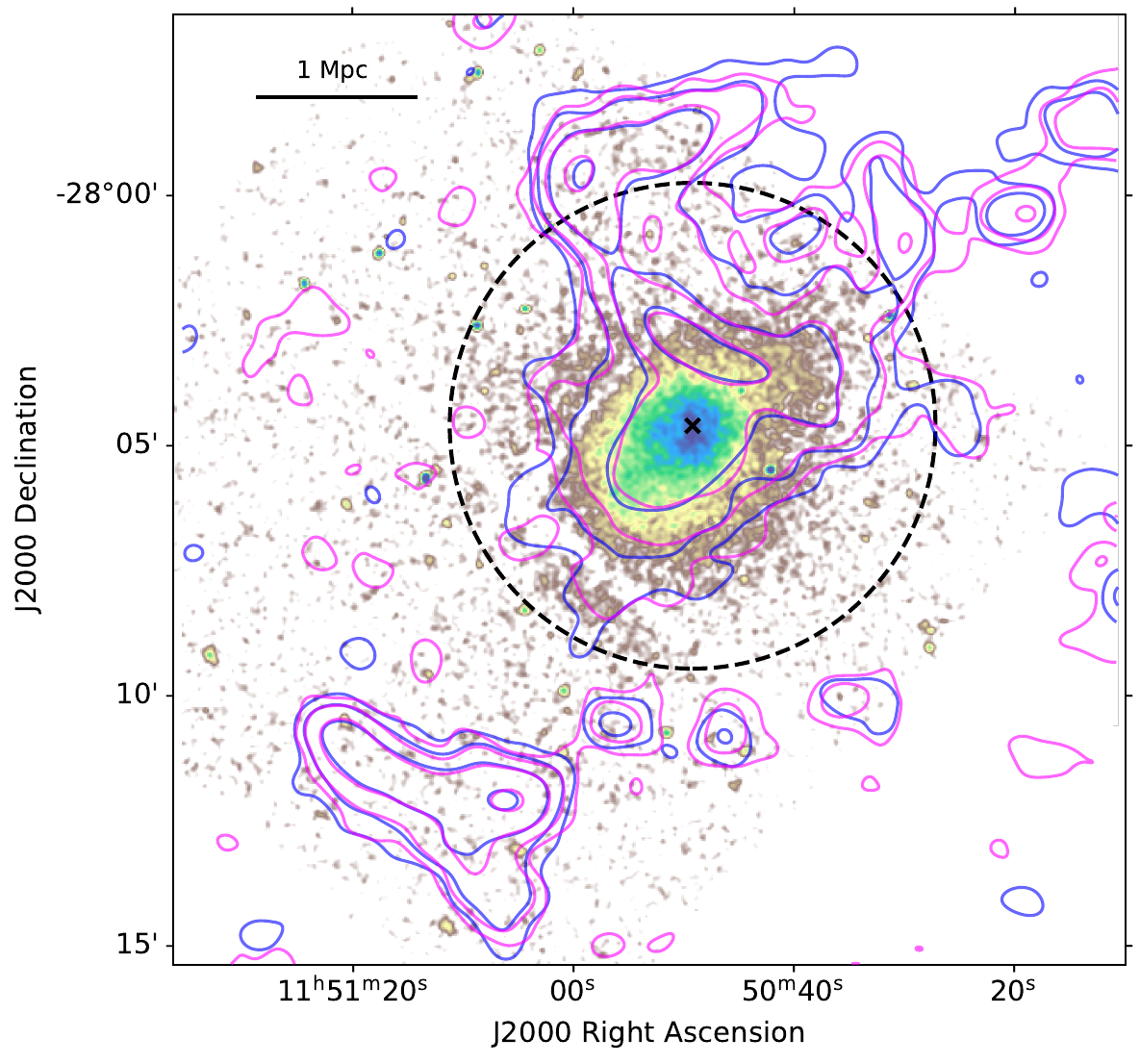}
    \caption{\textit{Chandra} image (in color) of PLCKG287 is shown here. The contours at [1,3,9,27...]$\times$ 3$\sigma_{\rm rms}$ from 40$''$ resolution images of uGMRT band 3 ($\sigma_{\rm rms}$ = 120 $\mu$Jy~beam$^{-1}$) and band 4 ($\sigma_{\rm rms}$ = 67 $\mu$Jy~beam$^{-1}$) are added in blue and magneta, respectively. The black `$\times$' sign indicates the peak of the X-ray emission and the black dashed circle indicates the R$_{500}$ of the cluster.}
    \label{X-ray_overlay}
\end{figure}

\subsection{Surface brightness profile} \label{subsec: surface_brightne}

We present the azimuthally averaged radio surface brightness profile for the central diffuse emission} (Figure~\ref{megahalo_SB}), following the procedure described in \citet{2009A&A...499..679M}. We used low-resolution (40$''$) images after subtracting the compact sources in the \textit{uv}-plane.  Due to their intervention in accurately estimating the surface brightness profile for the central diffuse radio emission, we have masked the emission from the NW relic, the filaments, and some residuals from bright diffuse sources. We also masked pixels below 1.5$\sigma_{\rm rms}$ to avoid regions dominated by noise and/or negative flux. Elliptical annuli have been used instead of circular annuli to account for the elongated morphology of the diffuse source. We have estimated the averaged radio brightness from each concentric elliptical annulus, centered on the peak of the radio halo emission, where the width of the annulus along semi-major axis has been chosen to be half of the synthesized beam of the image. We have considered only those annuli, where the averaged surface brightness is 3 times the background RMS noise.  

The azimuthally averaged radio surface brightness profile is shown in Figure~\ref{megahalo_SB}, where a discontinuity is seen in the profile for both frequencies. The discontinuity is seen around $\sim 800$ kpc at 400 MHz, and the emission before discontinuity is well-fitted with an exponential profile given by,

\begin{equation} \label{eq: elliptical_exp}
    I(r) = I_{0} \: e^{-\sqrt{\frac{x^{2}}{r_{1}^{2}} + \frac{y^{2}}{r_{2}^{2}}}}
\end{equation} 

where, $I_{0}$ is the central surface brightness and, $r_{1}$ and $r_{2}$ are e-folding radii along the semi-major and semi-minor axes, respectively. The fitting procedure is very similar to \citet{2022Natur.609..911C}, where a model (given in Equation~\ref{eq: elliptical_exp}) of the same pixel size of the radio image was created and then it was convolved with a Gaussian, having the full-width half-maximum (FWHM) equal to the synthesized beam of the radio image. Using the same annular region, the azimuthally averaged exponential model is derived. 

\begin{figure*}
\centering
    \includegraphics[width=0.46\textwidth]{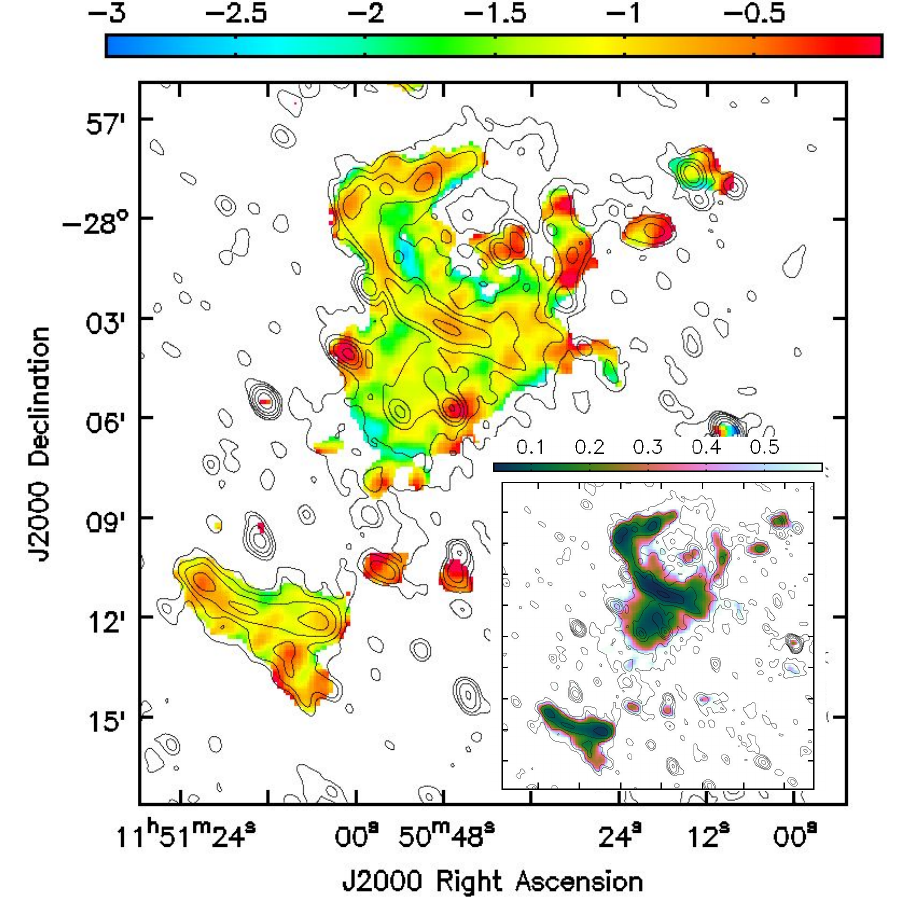}
    \includegraphics[width=0.48\textwidth]{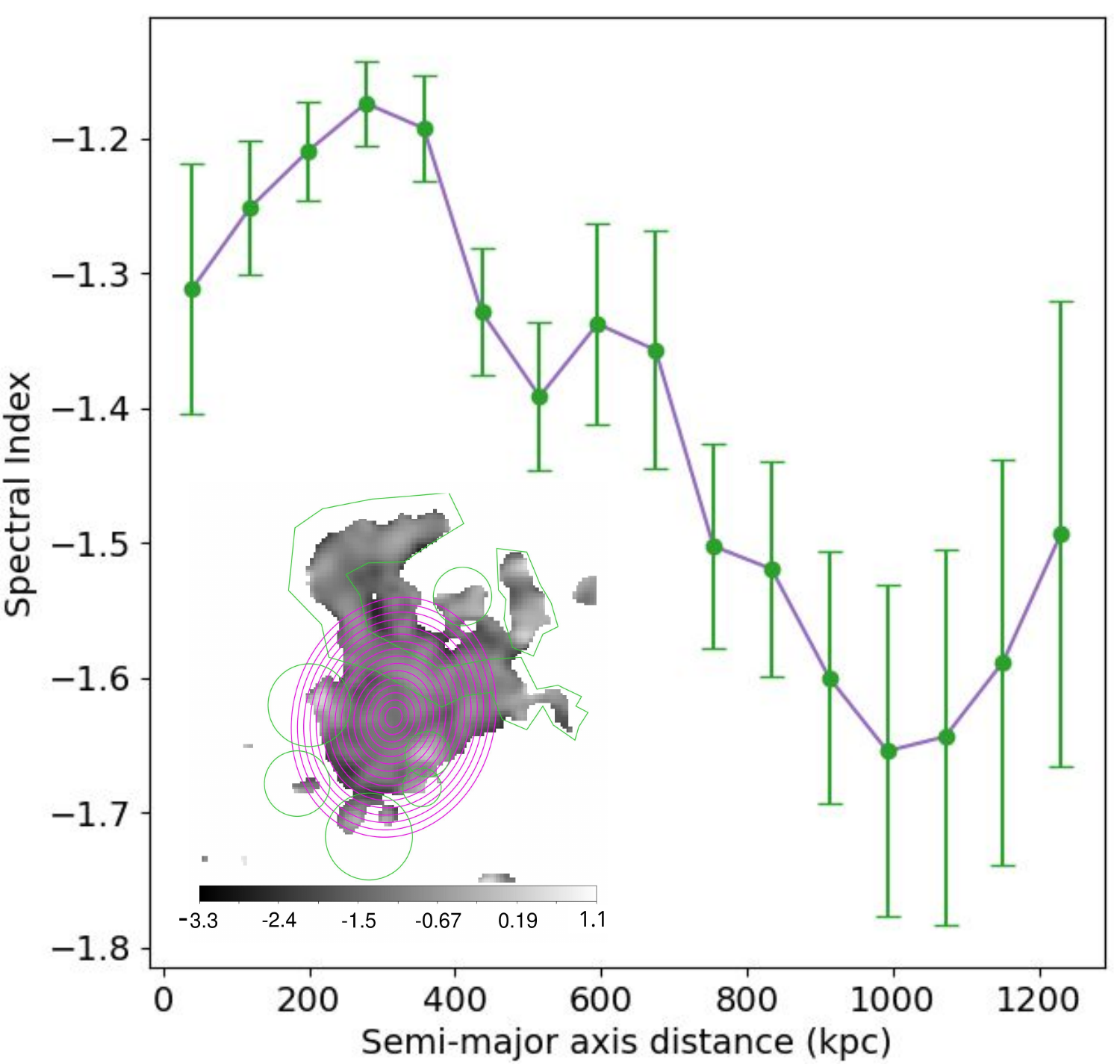}
     
    \caption{\textit{Left:} Spectral index map for the radio halo in PLCKG287 at 30$''$ is shown between 400 MHz and 700 MHz. The black contours are drawn at 3$\sigma_{\rm rms}$ $\times$ [1,2,4,8..] from the 400 MHz image. The corresponding error map is shown in the inset, and at the central region, overall the errors are \textless 0.2. \textit{Right:} The radial profile of the spectral index is shown in this panel. The outer component has a significantly steeper spectrum, compared to the radio halo. The annular region used to get spectral profile is shown in magenta color in the inset and masked areas are shown with green regions. }
    \label{Spidx_map}
\end{figure*}

The emission before the discontinuity is characterized as a radio halo, while the emission after the discontinuity is described as a megahalo. We point out that disentangling the contribution of the NW relic and filaments from the megahalo emission in the region above the NW relic is very difficult, as the relic does not show a clear flat edge (Section~\ref{subsec: sepctral_properties} and Figure~\ref{Spidx_map}) towards the outskirts. Therefore, we have decided to mask the northern part of the emission entirely and computed the surface brightness profile using only the southern region. 
This surface brightness profile confirms that the discontinuity in surface brightness remains present even when only the southern part is considered (Figure~\ref{megahalo_SB}), although the discontinuity is less prominent. These results reinforce the detection of the megahalo. The surface brightness of the radio megahalo is fainter by order of magnitude compared to the radio halo.

\subsection{Flux density measurements}

Point source subtracted images are used to calculate the integrated flux densities for radio halo (30$''$ resolution) and megahalo (40$''$ resolution). Integrated flux densities for radio halo are comparatively higher, $84.77\pm5.05$ mJy at 400 MHz and $41.75\pm2.56$ mJy at 700 MHz. The discrepancy ($\sim$ 34\% and 60\% at band 3 and 4 respectively) in the flux density measurements, compared to \cite{2014ApJ...785....1B}, is attributed to the wide bandwidth of the uGMRT compared to the legacy GMRT, which also enables the detection of the radio halo out to larger radii. Estimating the the flux density of the megahalo is difficult due to the presence of large-scale NW relic, radio halo, and several discrete sources. However, the flux density is calculated using the mean surface brightness in a region that excludes the central halos and NW relic and then multiplied by the total area within the 3$\sigma_{\rm rms}$ contours (similar method followed by \citealt{2022Natur.609..911C}). These flux density values for megahalo are $77.49\pm4.66$ mJy and $37.04\pm2.25$ mJy at 400 and 700 MHz, respectively. The uncertainties in the flux density ($S$) measurements were calculated using $\sigma_S = \sqrt{\sigma_{\rm cal}^2 +  \sigma_{\rm sub}^2 + N_{beam}\sigma_{\rm rms}^2}$ where, $\sigma_{\rm cal} = 0.05S$ represents error due to calibration uncertainties, $\sigma_{\rm sub} = 0.03S$ represents error due to point source subtraction and $N_{beam}$ is the number of beams across the diffuse emission. 

We have extrapolated these flux densities to 1.4 GHz using the spectral index of -1.27 for radio halo and -1.30 for radio megahalo (Section~\ref{subsec: sepctral_properties}) to estimate the radio powers and hence emissivities, and our values are in line with \citet{2022Natur.609..911C}. Estimated values for megahalo are upper limits as spectral index for it is $\leq$ -1.3. The calculated emissivities are volume-averaged, assuming the emission originates from an ellipsoid with semi-axes a, b, and c, where b=c. 

\begin{equation}
    J_{1.4GHz} = \frac{P_{1.4GHz}}{\frac{4}{3}\pi abc}
\end{equation}

The emissivity for radio megahalo is $ 3.58 \pm 0.22 \times 10^{-43}$ erg s$^{-1}$ cm$^{-3}$ Hz$^{-1}$ and for radio halo is $ 7.31 \pm 0.44 \times 10^{-42}$ erg s$^{-1}$ cm$^{-3}$ Hz$^{-1}$.  The megahalo is $\sim 20$ times fainter than the radio halo.

\begin{figure*} 
    \centering
    \includegraphics[width=0.90\textwidth]{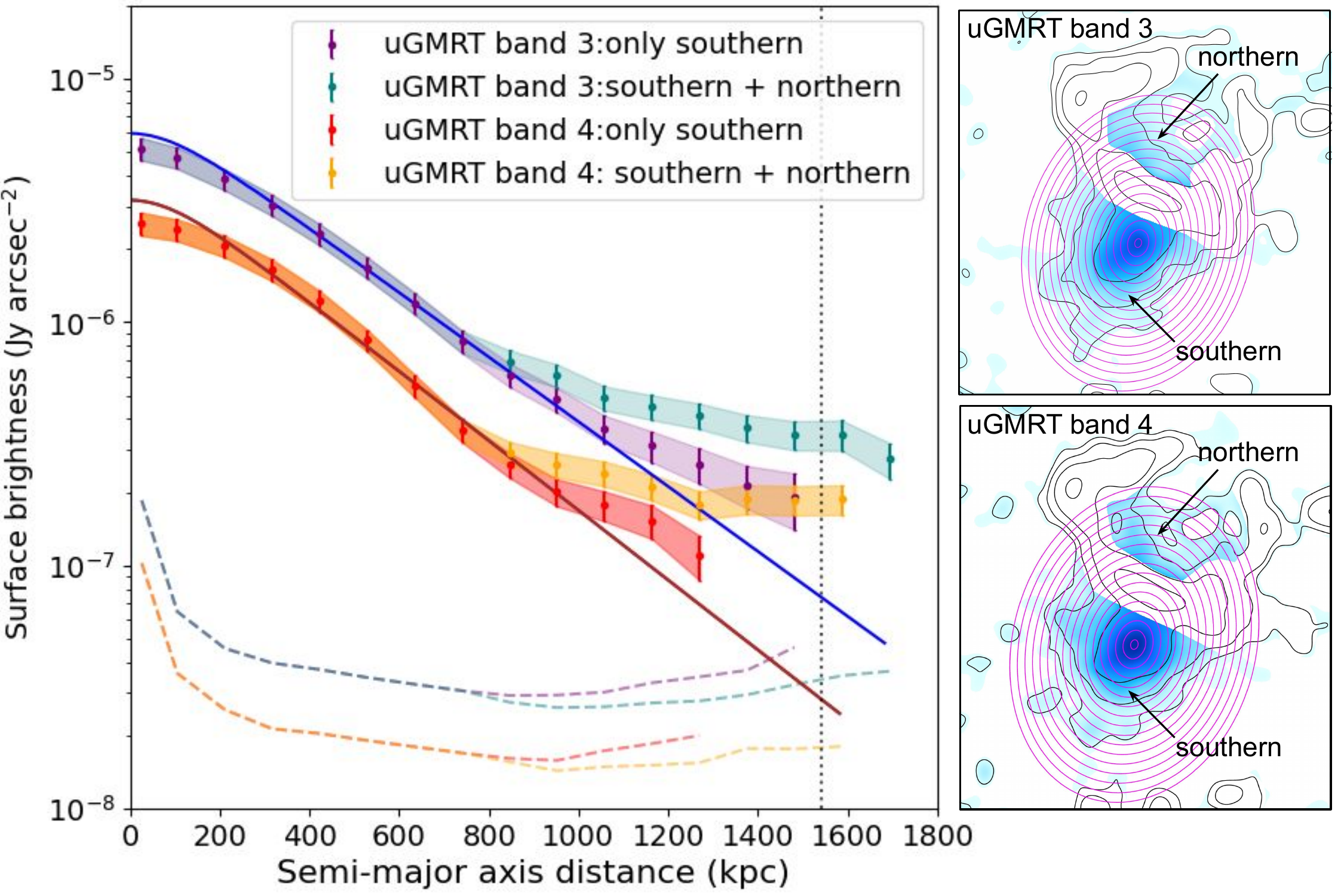}
    \caption{The surface brightness radial profile is shown, measured on the elliptical annular regions (inset images). The purple and red colors represent the profiles for only the southern diffuse emission to the NW relic in bands 3 and 4, respectively. The teal and orange colors represent the profiles considering both northern and southern diffuse emissions to the NW relic in Bands 3 and 4, respectively. The solid line is the exponential fit for the radio halo data. The dashed line marks the detection limit at 1$\sigma$ for each annulus, and the vertical dotted line shows the cluster's R$_{500}$. The annotated figures show the central diffuse emission in uGMRT bands 3 and 4, having a resolution of 40$''$, and RMS values of 120 and 67 $\mu$Jy~beam$^{-1}$, respectively. Contours are same as given in Figure~\ref{X-ray_overlay}. The region below 1.5$\sigma_{\rm rms}$ is masked while producing the surface brightness profile.}
    \label{megahalo_SB}
\end{figure*}

\subsection{Spectral properties} \label{subsec: sepctral_properties}

 Spatially-resolved spectral index maps are vital for understanding the CRe acceleration mechanism behind the radio halo emission \citep[e.g.,][]{2016ApJ...818..204V, doi:10.1126/sciadv.abq7623}. Hence, the spectral index map is created using the point source subtracted visibilities in band 3 and band 4, with a similar inner \texttt{uv} cut-off (\textgreater 0.1 k$\lambda$), and uniform weighting. Both the images are convolved to a 30$''$ resolution and the pixel values of less than 3$\sigma_{\rm rms}$ are blanked for each image.

The spectral index of the diffuse emission in PLCKG287 ranges from around -0.5 to -2.5 (Figure~\ref{Spidx_map}, left).  Two regions with very flat spectral index (\textgreater -0.5) are seen, which may be related to the residuals of the compact discrete sources.  The spectral index distribution across the radio halo is quite uniform. At the central region, the spectral index is comparatively flat ($\sim$ -1.2), whereas at the periphery (megahalo region) some patches of the ultra steep spectral index ($\lesssim$ -1.5) are seen, although this is the region with the largest uncertainties (inset of Figure~\ref{Spidx_map}, left). The integrated spectral index for radio halo region is ($\alpha_{400MHz}^{700MHz} = -1.27\pm0.22$) calculated using the flux densities given in Section~\ref{subsec: continuum_images}, in line with \citet{2014ApJ...785....1B}. Getting an integrated spectral index for the total megahalo emission is difficult mainly due to the presence of NW relic, radio halo as well as the contribution from compact sources. But from the spectral index map, the spectral index for megahalo is significantly steeper than the radio halo. The spectral index profile (Figure~\ref{Spidx_map}, right) also shows the sign of the distinction between the radio halo and the megahalo (although the errorbars are significantly high at the outskirts region), where the spectral index shows steepening beyond $\sim 750$ kpc, from around -1.3 to -1.7.

\section{Discussion} \label{sec:discussion}

The distinct diffuse emission surrounding the radio halo is confirmed to be a megahalo for the following reasons: i) the diffuse emission is detected up to $\sim R_{500}$ of the cluster, mainly at band 3, ii) the surface brightness profile shows the discontinuity at around $\sim 800$ kpc i.e. beyond radio halo of the cluster, iii) Spectral steepening is observed, as indicated by both the spectral map, which shows patches of steep spectral regions, and the spectral profile, which exhibits a tentative sign of steepening in the outskirts region, and iv) radio emissivity is much lower compared to radio halo ($\sim 20$ times). These are similar properties that are found for the megahalos discovered with LOFAR by \citet{2022Natur.609..911C}.

The origin of these megahalos is unclear. \citet{2022Natur.609..911C} found that a baseline turbulent energy flux originating from ongoing matter accretion can play a role behind the megahalo origin. \citet{2024ApJ...961...15N} recently showed that re-accelerating electrons from solenoidal turbulence at the cluster outskirts, with about $\sim$1\% efficiency, can account for the observed megahalo emission at such large scales. \citet{2024A&A...690A..67B} conducted a detailed numerical simulation of a single cluster, having very different mass compared to the clusters hosting megahalos and also compared to PLCKG287, to trace the origin of megahalo and found that Fermi II acceleration is responsible for the radio emission observed in galaxy clusters. Their study revealed that turbulence generates asymmetric, elongated diffuse radio emission after the latest major merger with properties similar to megahalos. The elongated shape and the steep spectrum of the megahalo in PLCKG287 are in line with this recent simulation. \citet{2014ApJ...785....1B} identified the SE-NW axis as the merging axis for the cluster PLCKG287 that experienced a two-stage merger, and the merger-driven turbulence may be able to accelerate the electrons in the outer region of the clusters. 
 
\section{Conclusions} \label{sec:summary}

We report here the uGMRT discovery of a radio megahalo in PLCKG287. The sensitive uGMRT observations enable us to map the low surface brightness emission extended up to the cluster outskirts. The new observations were three times more sensitive than the previously available low-frequency observations, signposting the detection of all known structures with high significance. The radio halo emission is detected up to 3.2 Mpc ($>$ R$_{500}$) and 2.9 Mpc at bands 3 and 4, respectively, with an elongated shape along the NW-SE axis. The azimuthally averaged radial surface brightness profile of the large-scale halo emission shows a flattening beyond $\sim$ 0.5 R$_{500}$, a property of a megahalo. We estimated the 1.4 GHz emissivity of the outer envelope to be $\sim 20$ times lower, and the spectral index to be steeper than the classical radio halo. The resolved spectral map shows uniformity over the extent of the halo, with some patches of the steep spectral regions ($\lesssim$ -1.5) at the periphery. The morphology of the PLCKG287 is elongated along the NW-SE axis and experienced a two-staged merger, in line with the recent hydrodynamical simulation, where the radio halo forms and gradually fades (indicated by a steepening of the spectral index) after the first merger, and a subsequent second merger develops the outer megahalo emission.

The detection of a radio megahalo at band 3 and band 4 of the uGMRT marks a significant step in our understanding of these large-scale structures. This discovery provides a more detailed view of the spectral properties and offers new insights into the underlying physical processes behind the megahalo origin. Due to the significantly steep spectrum of the megahalo emission, they were thought to have originated from a very non-efficient acceleration process, however, their detection at higher frequencies can help distinguish between different emission mechanisms at such a large distance from the cluster center. This detection not only highlights the capability of uGMRT to uncover such structures but also emphasizes that megahalos are not limited to observations at very low frequencies, such as those from LOFAR, and these large-scale phenomena might be more common in galaxy clusters than previously thought. As more radio megahalos are detected at frequencies beyond 144 MHz, they could provide key constraints on cluster dynamics, particle acceleration processes, and the cosmic-ray populations that permeate the vast intergalactic medium.

\begin{acknowledgments}

We thank the anonymous referee for their constructive comments that have improved the clarity of the paper. S.S., R.S., and R.K. acknowledge the support of the Department of Atomic Energy, Government of India, under project no. 12-R\&D-TFR-5.02-0700. R.S. thanks Virginia Cuciti for her quick replies to many useful queries. R.K.  acknowledges the support from the SERB Women Excellence Award WEA/2021/000008.
We thank the staff of the GMRT that made these observations possible. The GMRT is run by the National Centre for Radio Astrophysics (NCRA) of the Tata Institute of Fundamental Research (TIFR). The scientific results reported in this article are based in part on data obtained from the \textit{Chandra} Data Archive. This research had made use of the NASA/IPAC Extragalactic Database (NED), which is operated by the Jet Propulsion Laboratory, California Institute of Technology, under contract with the National Aeronautics and Space Administration. 

\end{acknowledgments}

\facilities{upgraded Giant Metrewave Radio Telescope (uGMRT), CXO}

\software{WSClean, CASA, SPAM, ds9}

\bibliography{PLCK287_ApJL_rev3}{}
\bibliographystyle{aasjournal}

\end{document}